\documentstyle[12pt]{article}
\newcommand{\be}{\begin{equation}}
\newcommand{\ee}{\end{equation}}
\newcommand{\nMean}{\langle n\rangle}

\begin{document}

\begin{center}

{\bf Moments of the truncated multiplicity distributions}\\

\vspace{2mm}

I.M. Dremin, V.A. Nechitailo\\

Lebedev Physical Institute, Moscow\\

\end{center}

\begin{abstract}
In experiment, the multiplicity distributions of inelastic processes are
truncated due to finite energy, insufficient statistics or special choice of
events. It is shown that the moments of such truncated multiplicity distributions
possess some typical features.In particular, the oscillations of cumulant moments 
at high ranks and their negative values at the second rank can be considered
as ones most indicative on specifics of these distributions. They allow to
distinguish between distributions of different type.

\end{abstract}

\section{Introduction}

Studies of multiplicity distributions of high-energy inelastic processes have
produced many important and sometimes unexpected results
(for the reviews, see, e.g., \cite{dre1, ddk, dgar}). The completely new
region of very high multiplicities will be opened with the advent of
RHIC, LHC and TESLA accelerators.

Theoretical approaches to multiplicity distributions in high-energy processes
have usually to deal with analytic expressions at (pre)asymptotic energies
which only approximately account for the energy-momentum conservation laws or
with purely phenomenological expressions of the probability theory. The
multiplicity range extends in this case from zero to infinity.

In experiment, however, one has to consider distributions truncated at some
multiplicity values in one or another way. These cuts could appear due to energy
limitations, low statistics of experimental data or because of
special conditions of an experiment. Energy limitations always impose the
upper cotoff on the tail of the multiplicity distributions. Low statistics of
data can truncate these distributions from both ends if it is
insufficient to detect rare events with very low and/or very high multiplicity.
Similar truncations appear in some specially designed experiments \cite{niki},
when events within some definite range of multiplicities have been chosen.

It would be desirable even in these cases
to compare the distributions within those limited regions with underlying
theoretical distributions. The straightforward fits are sometimes not accurate
enough to distinguish between various possibilities because the probability
values vary by many orders of the magnitude. More rigorous approach is
to compare different moments of the truncated distributions. Among them, the
cumulant moments $K_q$ seem to be most sensitive to slight variations (and,
especially, cuts and shoulders) of the distributions. They often reveal such
tiny details of the distributions which otherwise are hard to notice.

In particular, QCD predicts quite peculiar behaviour of cumulant moments as
functions of their rank $q$. According to solutions of the equations for the
generating functions of the multiplicity distributions in the asymptotic
energy region, the ratio of cumulant moments $K_q$ to factorial moments $F_q$
usually denoted as $H_q=K_q/F_q$ behaves as $q^{-2}$ and at preasymptotic energy
values reveals the minimum \cite{13} at $q\approx 5$
with subsequent oscillations at higher ranks \cite{41, lo2}. Such a behaviour
has been found in experiment at presently available energies \cite{dabg, sld}.
The solutions of the corresponding equations for the
fixed coupling QCD also indicate on similar oscillations \cite{dhwa}. At
asymptotics, the oscillations should disappear and $H_q$ becomes a smoothly
decreasing and positively definite function of $q$, as mentioned above.

Neither of the distributions of the probability theory possesses these
features. Among them, the negative binomial
distribution (NBD) happens to be one of the most successful ones in the
description of global features of the multiplicity distributions \cite{giov}.
Let us remind that the negative binomial distribution is defined as
\be
P_n=\frac {\Gamma (n+k)}{\Gamma (n+1)\Gamma (k)}a^n(1+a)^{-n-k}, \label{prob}
\ee
where $a=\langle n\rangle /k$, $\langle n\rangle $ is the mean multiplicity,
$k$ is an adjustable parameter, and the normalization condition reads
\be
\sum _{n=0}^{\infty }P_n=1. \label{nor1}
\ee 
Its generating function is
\be
G(z)=\sum _{n=0}^{\infty }P_n(1+z)^n=\left (1-\frac {z\langle n\rangle }{k}
\right )^{-k}.
\label{gene}
\ee
The integer rank factorial and cumulant moments, and their ratio are
\be
F_q=\frac {1}{\langle n\rangle ^q}\frac{d^qG(z)}{dz^q}\vert _{z=0}=
\frac {\Gamma (q+k)}{\Gamma (k)k^q},   \label{fact}
\ee
\be
K_q=\frac {1}{\langle n\rangle ^q}\frac{d^q\ln G(z)}{dz^q}\vert _{z=0}=
\frac {\Gamma (q)}{k^{q-1}},   \label{cumu}
\ee
\be
H_q=\frac {\Gamma (q)\Gamma (k+1)}{\Gamma (k+q)}.   \label{rati}
\ee
$H_q$-moments at the parameter $k=2$ behave as $2/q(q+1)$, i.e. with the
power-law decrease reminding at large $q$ that of QCD, however, with a different
weight factor. Therefore, at first sight, it could 
be considered as a reasonably good analytic model for asymptotic behaviour of
multiplicity distributions. It has been proclaimed \cite{uglu, levc, glug, ugio}
that the superposition of two NBDs with different parameters and their cutoff
at high multiplicities can give rise to oscillations of $H_q$ and better fits
of experimental data at preasymptotic energies. Nevertheless, the fits have not
been perfect enough.

Let us compare first the asymptotic QCD predictions with NBD fits at different
values of the adjustable parameter $k$. The values of $D_q=q^2H_q$ are plotted
in Fig. 1 as functions of $q$ for the asymptotic QCD (where they are identically
equal to 1) and for NBD at different $k$. At $k=2$, they exceed 1 tending to 2
at large $q$. At larger values of $k$, all $D_q$ are less than 1 except $D_2=1$
at $k=3$. Surely, the identity $D_1\equiv 1$ is valid for any $k$ due to the
normalization condition.

To get asymptotic QCD results with all $D_q\equiv 1$
from the expressions similar to NBD, one would need to modify NBD in such a
way that the parameter $k$ becomes a function of $n$. Thus some effective values
of $k$ should be used to get QCD moments $D_q=1$ at various $q$. These effective
values are shown in Fig. 1 at the corresponding QCD points. They are
obtained as the solutions of the equation
\be
\prod  _{n=1}^{q-1}\left (1+\frac {k(q)}{n}\right )=q^2,      \label{piq2}
\ee
which follows from Eq. (\ref{rati}) for $H_q=q^{-2}$. They show that
$k$ somewhat decreases from 3 to some values exceeding 2 with increase of $q$.
This reflects the well known fact that the tails of distributions are 
underestimated in NBD-fits \cite{glug} compared to experimental data in the
preasymptotic region. Also, the amplitude of oscillations and their
periodicity are not well reproduced by a single truncated NBD \cite{glug},
and one has to use the sum of at least two NBDs to get a better fit.
However, rather large values of $k$ were obtained in these fits. It implies,
in fact, that the fit is done with the help of two distributions very close
to Poissonian shapes because the Poisson distribution is obtained from
NBD in the limit $k\rightarrow \infty$. Therefore, the tails are suppressed
very strongly.

Here, we will not try to fit experimental data focusing our efforts on 
qualitative changes of moments when NBD is truncated, especially, as applied
to studies of very high multiplicities.

In QCD considerations based on the equations for the generating functions for
quark and gluon jets, the preasymptotic (next-to leading order etc) corrections
give rise to oscillations of $H_q$. Even though they are of the higher order in
the coupling strength, they appear mainly due to account of energy
conservation in the vertices of Feynman diagrams but not due to considering
the higher order diagrams which are summed in the modified perturbation
theory series (see \cite{dgar}). In the phenomenological approach, this would
effectively correspond to the cutoff of the multiplicity distribution at
some large multiplicity. Therefore, we intend here to study how strongly
such a cutoff influences the NBD-moments, whether it produces oscillations of
the cumulant moments, how strong they are, and, as a more general case,
consider the moments of NBD truncated both at low and high multiplicities. This
would help answer the question if the shape of the distribution in the
limited region can be accurately restored from the behaviour of its moments.
It could become especially helpful if only events with very high multiplicities
are considered in a given experiment because of the above mentioned
underestimation of tails in the NBD-fits.

\section{Truncated NBD and its moments}

In real situations, the multiplicity distribution is sometimes measured in
some interval of multiplicities and one can try to fit by NBD the data
available only in the restricted multiplicity range. Therefore,
we shall consider the negative binomial distribution within the interval of
multiplicities $m\leq n\leq N$ called $P_n^{(c)}$ and normalized to 1 so that
\be
\sum _{n=m}^NP_n^{(c)}=1.   \label{norm}
\ee
Moreover, due to above reasoning and to simplify formulas we consider here
only the case of $k=2$. The generalization to arbitrary values of $k$ is
straightforward.

The generating function of the truncated distribution
$G_c(z)$ can be easily found as
\be
G_c(z)=\sum _{n=m}^NP_n^{(c)}(1+z)^n=
G(z)(1+z)^m \frac{f(z)}{f(0)}
,   \label{gcz}
\ee
where
\be
f(z)=1+m(1-x)-[1+(N+1)(1-x)]x^{N-m+1}
,    \label{fz}
\ee
\be
x=b(1+z), \;\;\;\; b=\frac {a}{1+a}.     \label{xb}
\ee
Correspondingly,
\be
f(0)\equiv f(z=0)\equiv f(x=b).    \label{fzxb}
\ee

Using the above formulas for the factorial moments, one gets the following
formula for the moments of the truncated distribution expressed in terms
of the NBD-moments (\ref{fact}):
\begin{eqnarray}
\langle n\rangle _c^qF_q^{(c)}=\sum _{r=0}^q\frac {\Gamma (q+1)}{\Gamma (r+1)
\Gamma (q-r+1)}\langle n\rangle ^{q-r}F_{q-r}\cdot \nonumber \\
\frac {\frac {\Gamma (m+2)}
{\Gamma (m+2-r)}[m(1-b)+1-r]-\frac {\Gamma (N+3)}{\Gamma (N+3-r)}b^{N-m+1}
[(N+1)(1-b)+1-r]}{m(1-b)+1-b^{N-m+1}[(N+1)(1-b)+1]},   \label{fcut}
\end{eqnarray}
where $\langle n\rangle _c$ is the mean multiplicity of the truncated 
distribution. It is related to the mean multiplicity $\langle n\rangle $ of
the original distribution as
\be
\langle n\rangle - \langle n\rangle _c =\frac {(1-b)[(N+1)(N+2)b^{N-m+1}-m(m+1)]}
{1+m(1-b)+b^{N-m+1}[(N+1)b-N-2]}.   \label{ncut}
\ee
Inserting formula (\ref{fact}) in (\ref{fcut}), one gets
\begin{eqnarray}
\langle n\rangle _c^qF_q^{(c)}=\sum _{r=0}^q\frac {(q-r+1)\Gamma (q+1)}
{2^{q-r}\Gamma (r+1)}\langle n\rangle ^{q-r}\cdot \nonumber \\
\frac {\frac {\Gamma (m+2)}
{\Gamma (m+2-r)}[m(1-b)+1-r]-\frac {\Gamma (N+3)}{\Gamma (N+3-r)}b^{N-m+1}
[(N+1)(1-b)+1-r]}{m(1-b)+1-b^{N-m+1}[(N+1)(1-b)+1]}.   \label{fcu2}
\end{eqnarray}
For computing it is more convenient to use the formula (\ref{fcu2})
in the following form:

$$
F^{(c)}_q = \left(\frac{\nMean}{\nMean_c}\right)^q F_q
\left\{
1+ \frac{1}{f(0)(q+1)}\sum^q_{r=1}\frac{a^{-r}}{r!}(q+1-r)\times
   \right.
$$

$$
[ ( \frac{m}{1+a}+1-r)\theta(m+1-r)\prod^r_{i=1}(m+2-i)
$$

\begin{equation}
\left.
-(\frac{N+1}{a+1}+1-r) b^{N-m+1} \prod^{r}_{i=1}(N+3-i) ]
\right\}        \label{fcom}
\end{equation}

These expressions can be used also for the distributions truncated at one side
by setting $m=0$ or $N=\infty $.

The cumulant moments can be calculated after the factorial moments are known
from Eq. (\ref{fcut}) according to the identities
\be
F_q=\sum _{m=0}^{q-1}\frac {\Gamma (q)}{\Gamma (m+1)\Gamma (q-m)}K_{q-m}F_m.
\label{fqkq}
\ee
This formula is a simple relation between the derivatives of a function and
of its logarithm (see Eqs (\ref{fact}) and (\ref{cumu})). Therefore it is valid
for both original and truncated distributions.

For the Poisson distribution, the ratios $H_q$ are identically equal to zero,
and are given by Eq. (\ref{rati}) for NBD while truncation induces new features.
In Figures we show the behaviour of the ratios $H_q$ as functions of
the rank $q$ for the truncated Poisson and negative binomial distributions.

At the beginning, we consider the abrupt cutoff only of the very high
multiplicity tail, i.e., the case $m=0$ and $N>\langle n\rangle $. This mimics
the energy-momentum conservation limits. In Figs 2 and 3, it is shown that
such a cutoff $N$ induces oscillations of $H_q$. The farther is the cutoff from
the mean multiplicity, the weaker are oscillations. This quite expected result
is known from long ago \cite{uglu, levc}. It
is demonstrated in Fig. 2 for $\langle n\rangle =10$ and different cutoffs
at $N=30, 40, 50$. Another representation of the same result is seen in Fig. 3
where the constant cutoff $N=30$ has been chosen for different $\langle n\rangle $
equal to 5, 10 and 15. The closer is the cutoff to $\langle n\rangle $, the
stronger the low-rank moments are damped. For the faraway cutoff $N=50$, the
period of oscillations increases. This increase is larger for lower mean
multiplicity (see Fig. 4). At $N/\langle n\rangle $=const, one
observes the approximate scaling of $H_q$ as seen in Fig. 5.

\section{Very high multiplicities}

With the advent of RHIC, LHC and TESLA we are approaching the situation when
average multiplicities become very high and the tails of multiplicity
distributions reach the values which are extremely large. These events with
extremely high multiplicities at the tail of the distribution can be of a
special interest. The tails of particular channels die out usually very fast,
and a single channel dominates at the very tail of the distribution. Mostly
soft particles are created in there. Thus one hopes to get the direct access to
very low-$x$ physics. QCD-interpretation in terms of BFKL-equation (or its
generalization) can be attempted. Also, the hadronic densities are rather high
in such events, and the thermodynamical approach can be applied \cite{sman}.

However, these events are rather rare and the experimental statistics is quite
poor until now. The Poisson distribution has the tail which decreases mainly
like an inverse factorial. According to NBD (\ref{prob}), the tail is
exponentially damped with the power-increasing preexponential factor. At the
same time, QCD predicts even somewhat slower decrease as is seen in Fig. 1
from the behaviour of moments. This is important for future experiments in
the very high multiplicity region.

To study these events within the truncated NBD with $k=2$ according to
Eqs (\ref{fcu2}), (\ref{fcom}), let us choose the multiplicity interval of the constant
length $N-m=20$ and place it at various distances from the mean multiplicity
$\langle n\rangle =10 $ as in Fig. 2. The resulting $H_q$ are shown in Fig. 6.
The most dramatic feature is the negative values of $H_2$ and the subsequent
change of sign of $H_q$ at each $q$ in the case when the lower cutoff $m$ is
noticeably larger than $\langle n\rangle $ ($m/\langle n\rangle \geq 2$).
This reminds of the behaviour of $H_q$ for the fixed multiplicity distribution
and shows that the NBD-tail decreases quite fast so that the multiplicity $m$
dominates in the moments of these truncated distributions. 

The same features are demonstrated in Figs 7 and 8 for different average
multiplicities and different positions of the fixed window $N-m=20$. 
In Fig. 7, the window is rather close to $\langle n\rangle $ or even contains
it inside (for $\langle n\rangle $=15). Therefore $H_2<0$ only for
$\langle n\rangle $=5. In Fig. 8, it is very far from $\langle n\rangle $.
Thus all $H_2$ are negative, the more the lower is $\langle n\rangle $.
Again, the sign-changing characteristics remind those for the fixed multiplicity
distribution.

Another possibility to study the tail of the distribution with the help of
$H_q$-ratios is their variation with the varying length of the tail chosen.
At the same mean multiplicity $\langle n\rangle $=10, we calculate moments for 
the intervals starting at $m=20$ and ending at $N=50,\; 40,\; 30,\; 25$. The
values of $H_q$ at rather low ranks $q=2,\; 3,\; 4,\; 5$ are very sensitive to
the interval length as shown in Fig. 9. The values of $H_2$ vary, e.g., by
the order of magnitude.

\section{Conclusions}

In connection with some experiments planned, our main concern here was to 
learn if $H_q$-ratios can be used to judge about the behaviour of the tail
of the multiplicity distribution. Using NBD as an example, we have shown that
$H_q$ behave in a definite way depending on the size of the multiplicity
interval chosen and on its location. Comparing the corresponding
experimental results with NBD-predictions, one would be able to show whether
the experimental distribution decrease slower (as predicted by QCD) or faster
than NBD. 

In particular, the negative values of $H_2$ noted above are of special interest
because they show directly how strong is the decrease of the tail. NBDs at
different $k$ values would predict different variations of $H_2$ with more
negative $H_2$ for larger $k$. Also, the nature of oscillations of $H_q$-moments
at larger values of $q$ reveals how steeply the tail drops down. 

Let us stress that the choice of high multiplicities for such a conclusion
could be better than the simpleminded fit of the whole distribution. As one
hopes, in this case there is less transitions between different channels of
the reaction (e.g., from jets with light quarks to heavy quarks), and the
underlying low-$x$ dynamics can be revealed.

{\bf Acknowledgements}

This work is supported by the RFBR grants N 00-02-16101 and 02-02-16779.\\

{\bf Figure captions.}\\

Fig. 1. $D_q=q^2H_q$ are shown for asymptotical QCD by circles 
($D_q^{QCD}\equiv 1$) in 
comparison with NBD-predictions at different values of the parameter $k$=2
(diamonds), 3 (crosses), 10 (squares). The numbers near the QCD values show
which values of $k$ one would need to use for NBD to fit $D_q=1$ at the 
corresponding $q$.

Fig. 2. $H_q$-moments for NBD with $\langle n\rangle $=10 truncated at
$N=30$ (diamonds), 40 (crosses), 50 (squares) at $m=0$.

Fig. 3. $H_q$-moments for NBD with $\langle n\rangle $=5 (diamonds), 10
(crosses), 15 (squares) truncated at $N=30$ at $m=0$.

Fig. 4. $H_q$-moments for NBD with $\langle n\rangle $=5 (diamonds), 10
(crosses), 15 (squares) truncated at $N=50$ at $m=0$.

Fig. 5. Approximate scaling of $H_q$-moments for NBD with
$N/\langle n\rangle $=const.

Fig. 6. $H_q$-moments for NBD with $\langle n\rangle $=10 truncated at
$m=10, N=30$ (diamonds), 20-40 (crosses), 30-50 (squares). 

Fig. 7. $H_q$-moments for NBD with $\langle n\rangle $=5 (diamonds), 10
(crosses), 15 (squares) truncated at $m=10, N=30$.

Fig. 8. $H_q$-moments for NBD with $\langle n\rangle $=5 (diamonds), 10
(crosses), 15 (squares) truncated at $m=30, N=50$.

Fig. 9. $H_q$-moments for NBD with $\langle n\rangle $=10 truncated at
$m=20, N=50$ (diamonds), 20-40 (crosses), 20-30 (squares), 20-25 (circles).

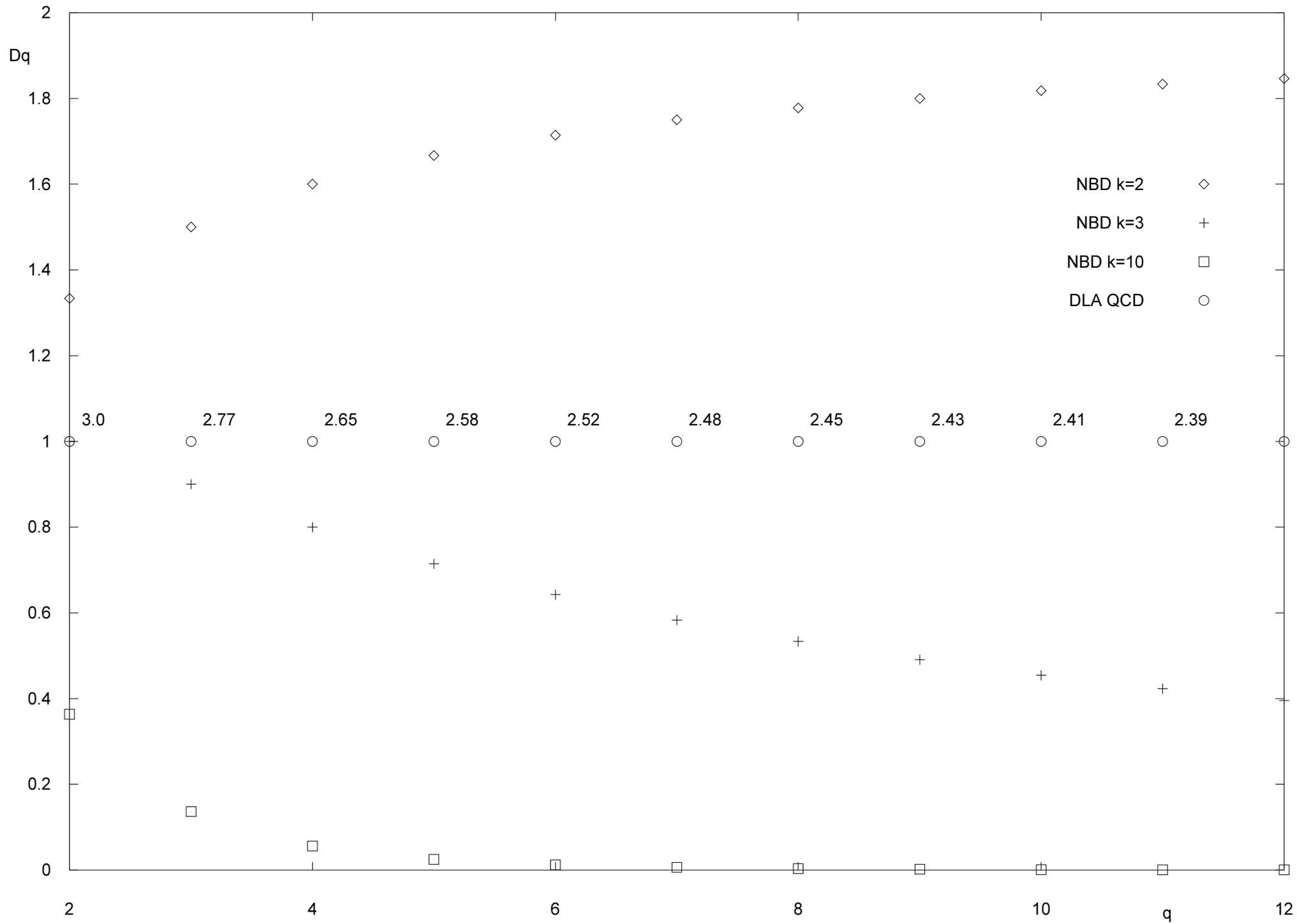
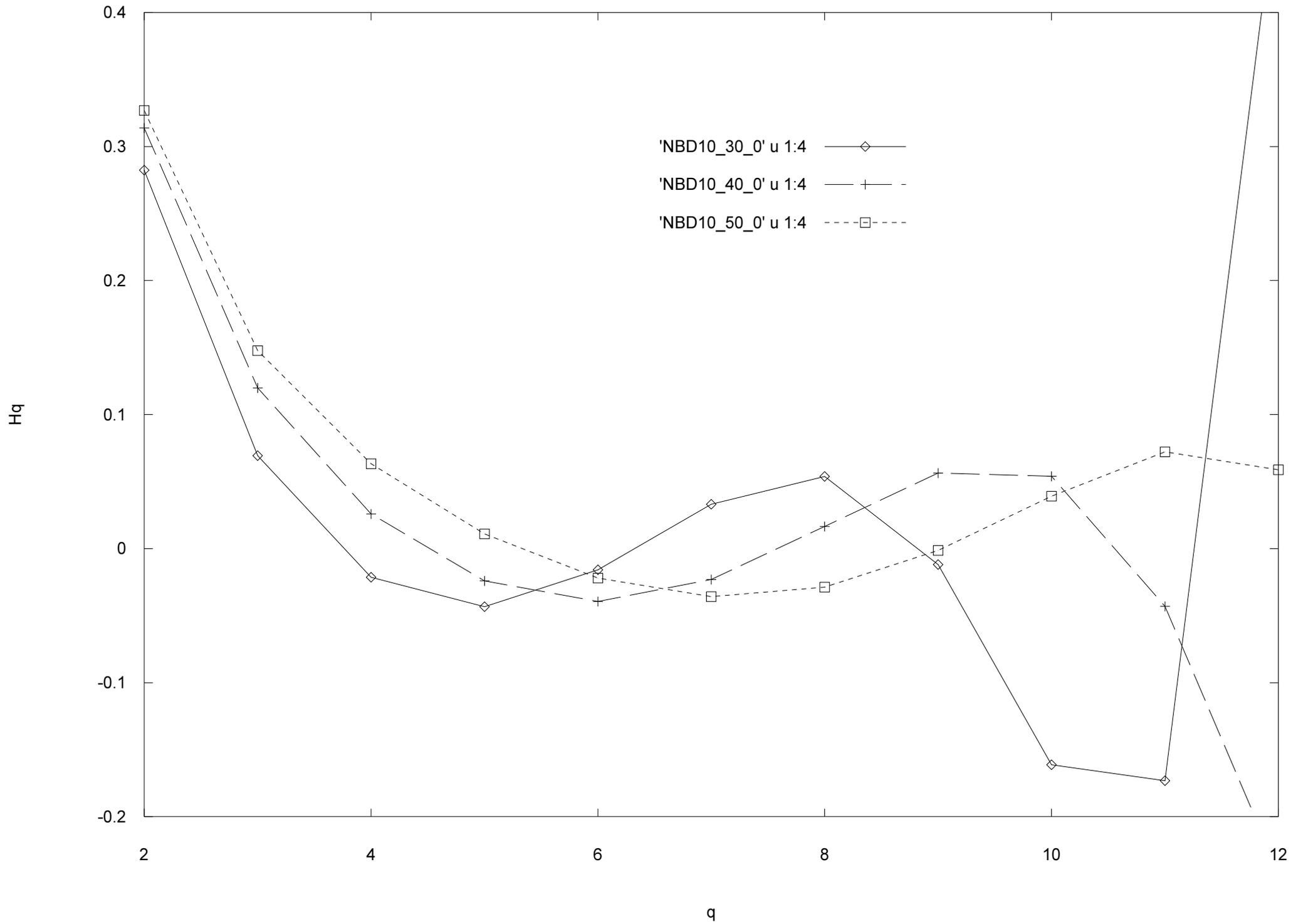
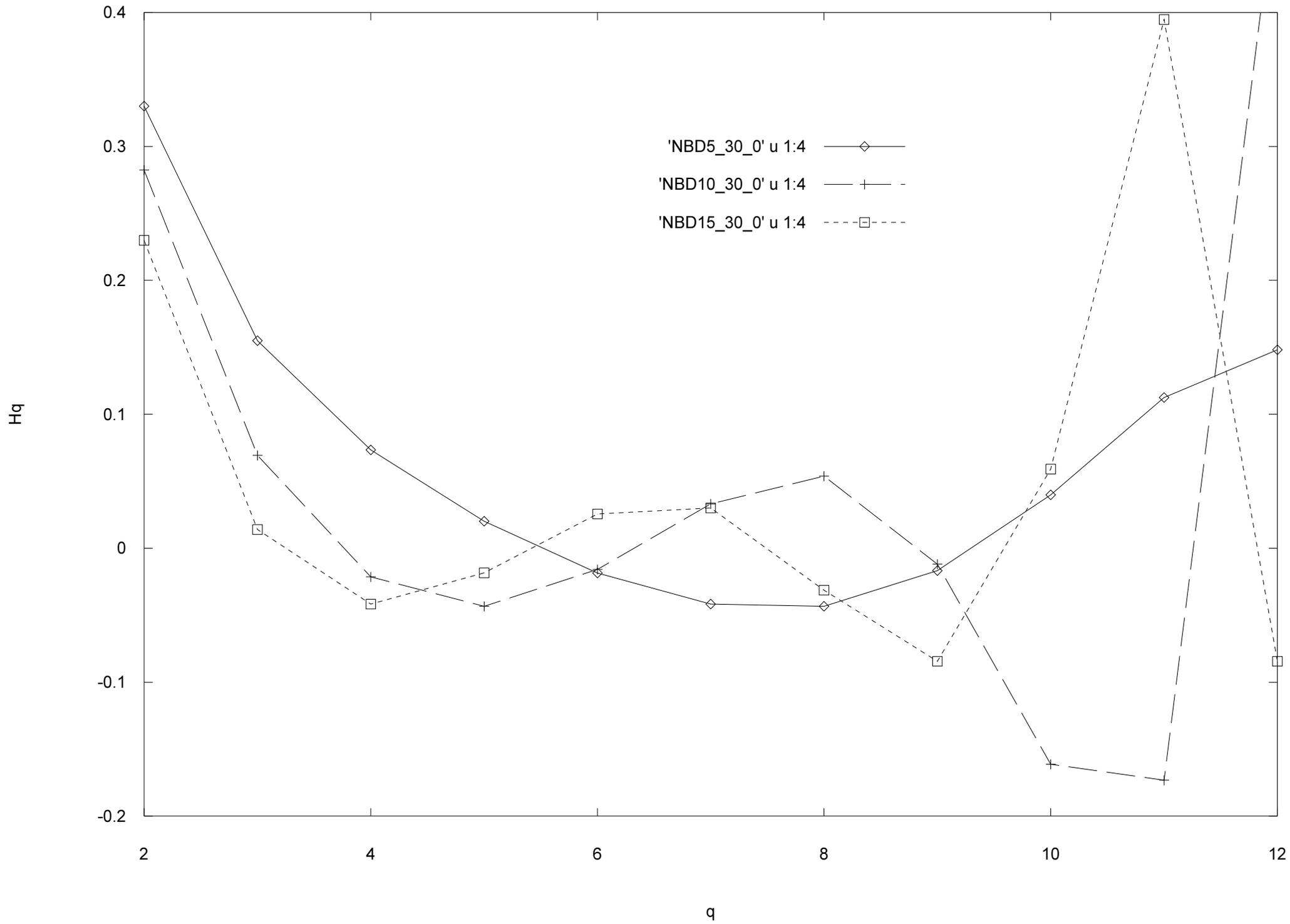
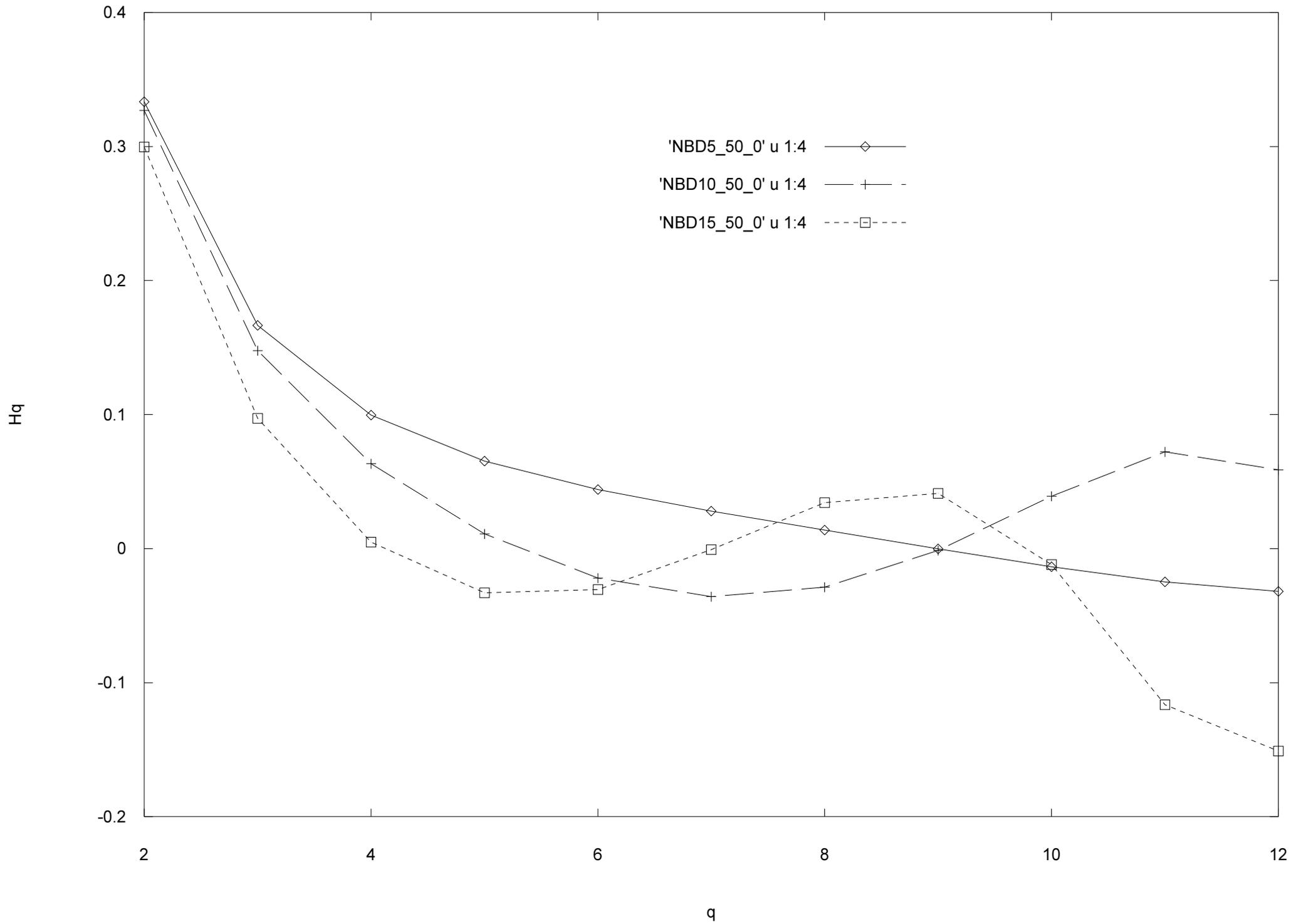
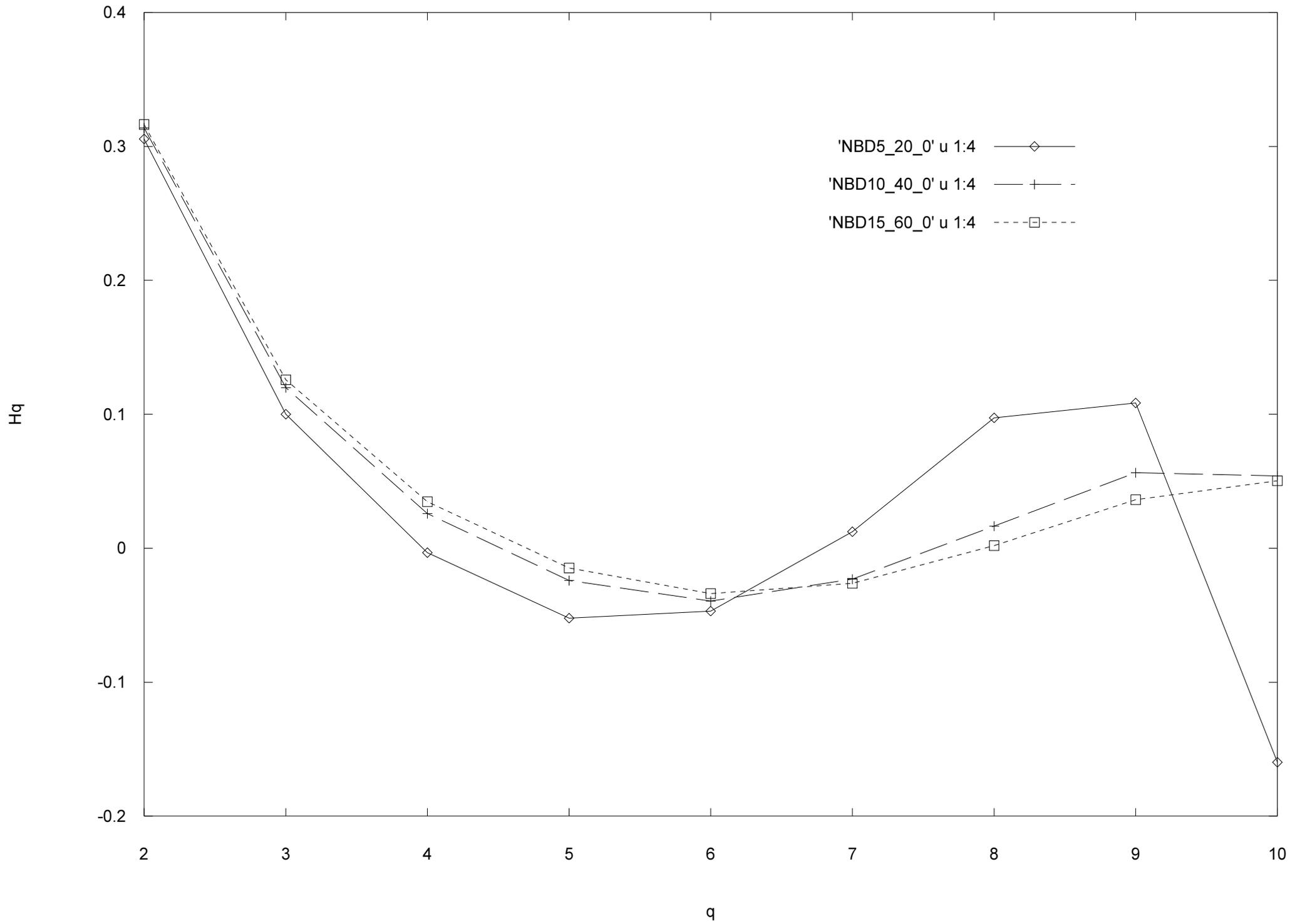
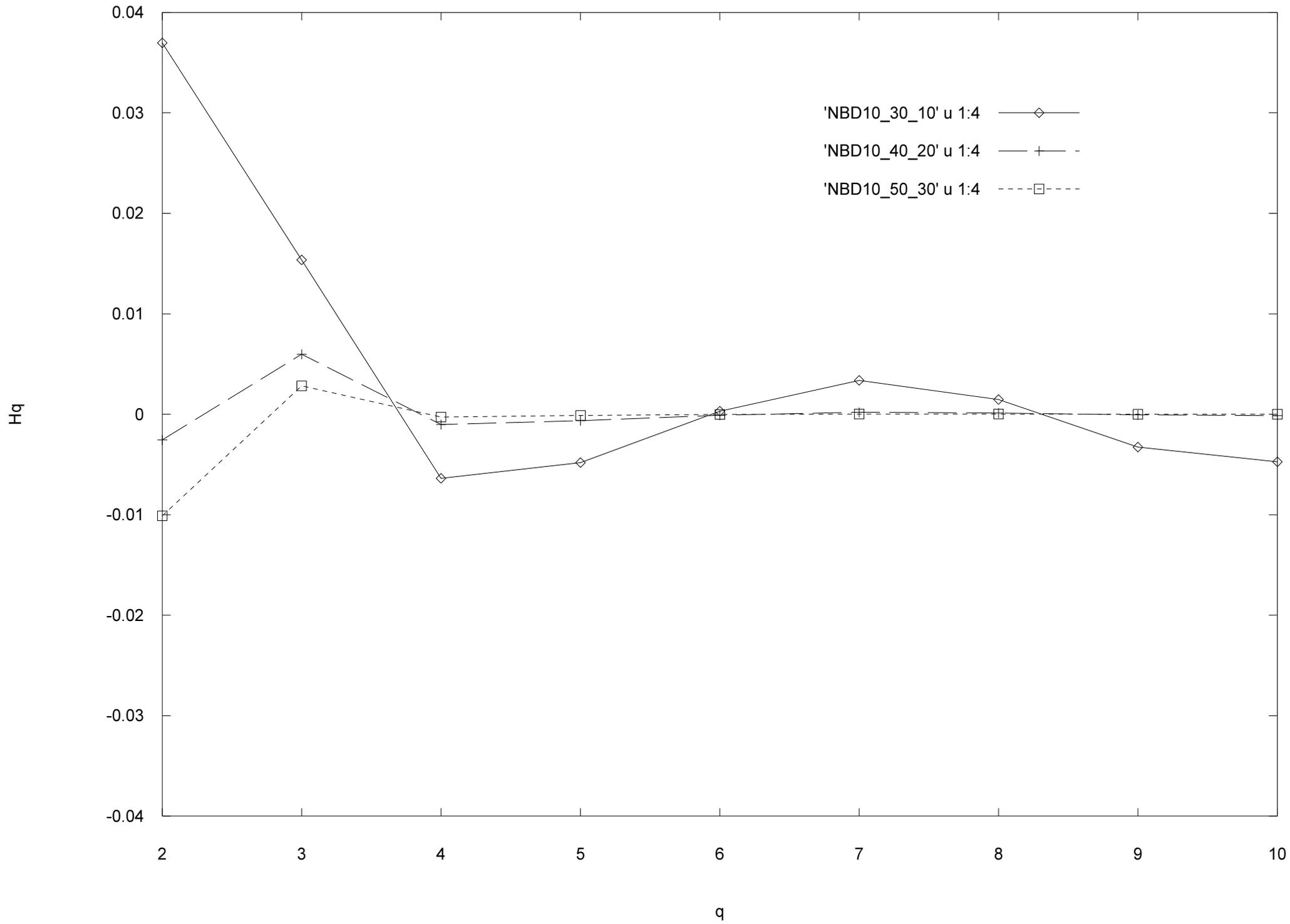
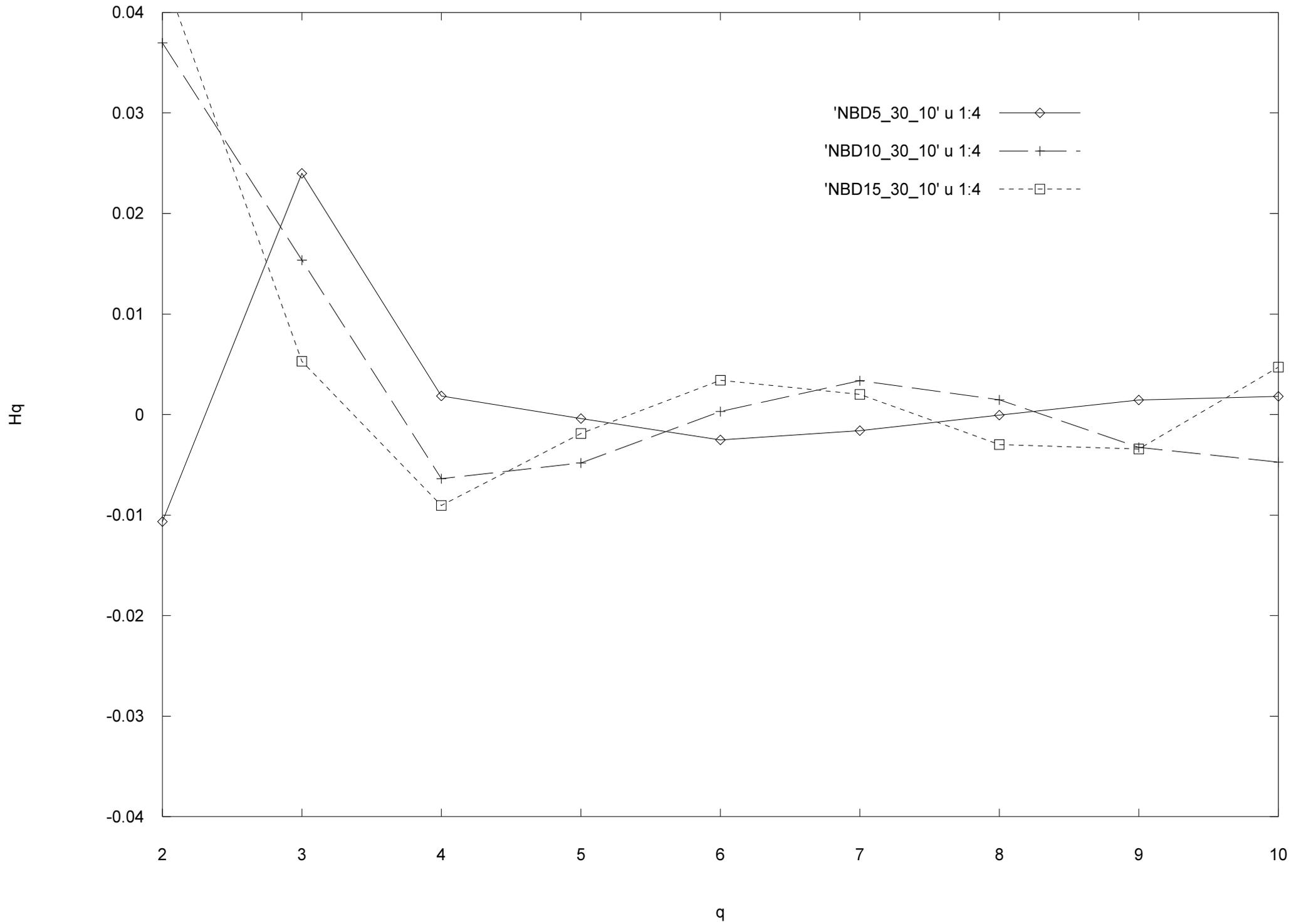
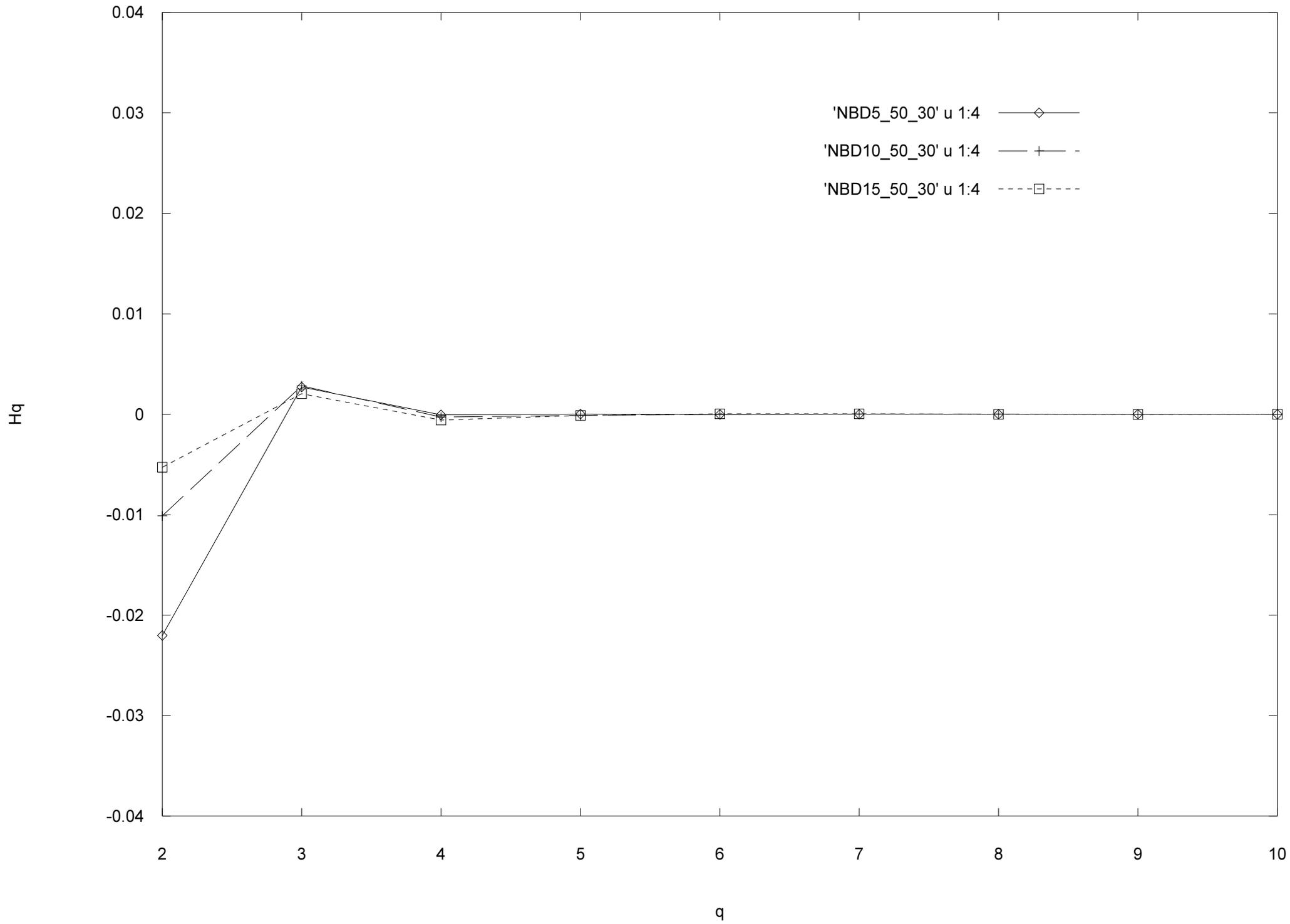
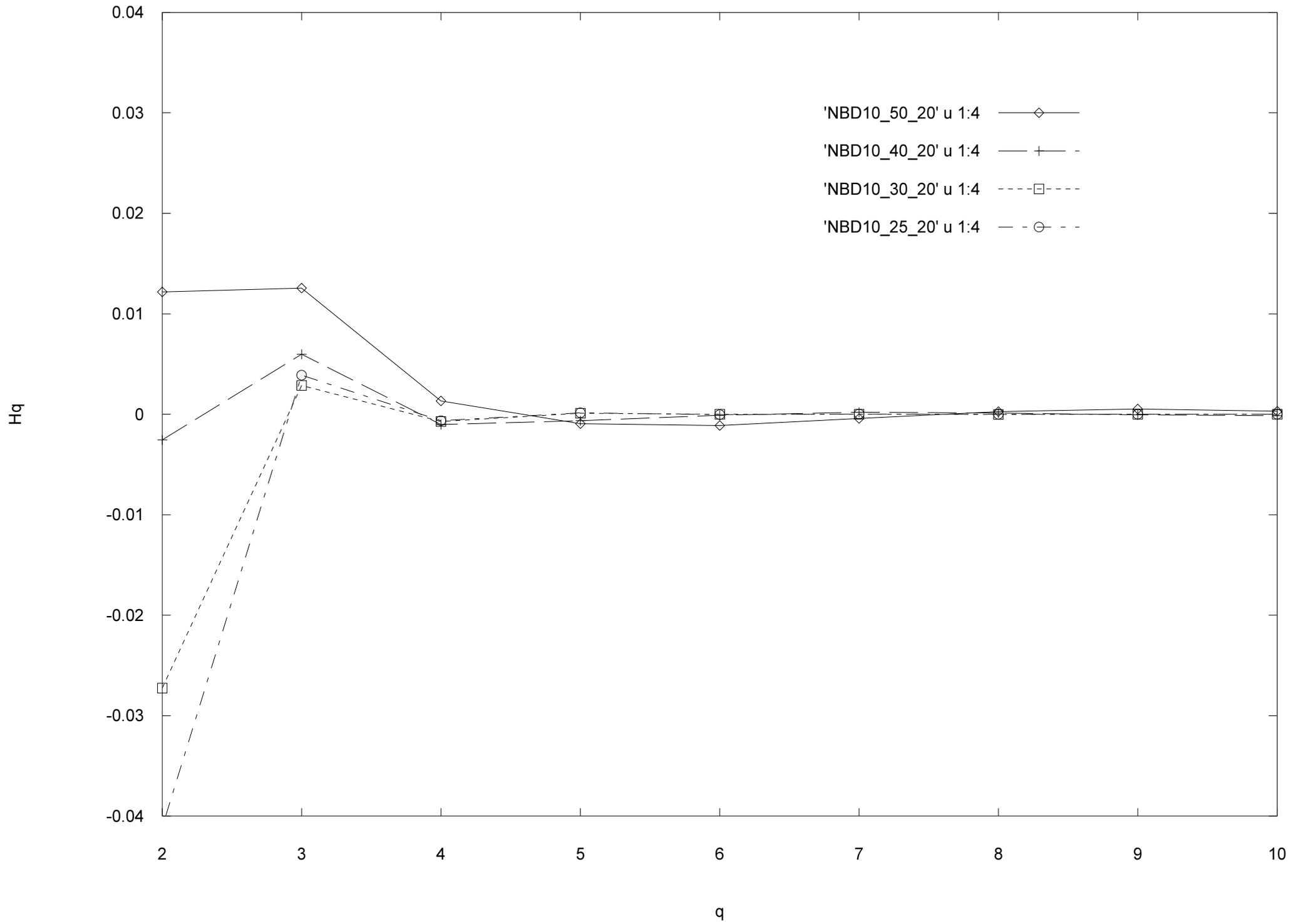
\end{document}